\documentclass[onecolumn,notitlepage,superscriptaddress,nofootinbib]{revtex4-1}
\usepackage{graphicx}
\usepackage{amsmath,amssymb,latexsym}
\usepackage{bm}
\usepackage{epsfig}

\newcommand{\eps}{\delta v}
\newcommand{\rar}{\rightarrow}

\newcommand{\half}{\frac{1}{2}}

\newcommand{\barr}{\begin{array}}
\newcommand{\earr}{\end{array}}
\newcommand{\bea}[1]{\begin{eqnarray} \label{(#1)}}
\newcommand{\eea}{\end{eqnarray}}
\newcommand{\beq}[1]{\begin{equation} \label{(#1)}}
\newcommand{\eeq}{\end{equation}}

\newcommand{\rf}[1]{(\ref{(#1)})}

\newcommand{\nutau}{\nu_\tau}
\newcommand{\nue}{\nu_e}
\newcommand{\numu}{\nu_\mu}

\newcommand{\nua}{\nu_{\alpha}}
\newcommand{\nub}{\nu_{\beta}}
\newcommand{\Uaj}{U_{\alpha j}}
\newcommand{\Ubj}{U_{\beta j}}

\newcommand{\tU}{{\tilde U}}
\newcommand{\tildeUaj}{{\tilde U}_{\alpha j}}
\newcommand{\tildeUbj}{{\tilde U}_{\beta j}}
\newcommand{\tildeUak}{{\tilde U}_{\alpha k}}
\newcommand{\tildeUbk}{{\tilde U}_{\beta k}}
\newcommand{\ta}{t_{\alpha}}
\newcommand{\tb}{t_{\beta}}
\newcommand{\va}{v_{\alpha}}
\newcommand{\vb}{v_{\beta}}
\newcommand{\thtilde}{{\tilde \theta}}
\newcommand{\Eres}{E_{\rm R}}
\newcommand{\Lres}{L_{{\rm R}(n)}}
\newcommand{\Dlsnd}{\Delta}
\newcommand{\Datm}{\Delta_{\rm atm}}

\begin{document}

\title{A model of superluminal neutrinos}

\author{D.~Marfatia}
\affiliation{
Department of Physics \& Astronomy, University of
Kansas, Lawrence, KS 66045, USA}

\author{H.~P\"as}
\affiliation{
Fakult\"at f\"ur Physik, Technische Universit\"at Dortmund,
D-44221 Dortmund, Germany}

\author{S.~Pakvasa}
\affiliation{Department of Physics \& Astronomy, 
University of Hawaii, Honolulu, HI 96822, USA}

\author{T.~J.~Weiler}
\affiliation{Department of Physics \& Astronomy, 
Vanderbilt University, Nashville, TN 37235, USA}

\begin{abstract}
  Motivated by the tentative observation of superluminal neutrinos by the OPERA experiment,
  we present a model of active-sterile neutrino oscillations in which sterile neutrinos are superluminal and active
  neutrinos appear superluminal by virtue of neutrino mixing.
  The model demonstrates some interesting possibilities and challenges 
  that apply to a large class of models aiming to explain the OPERA result.
\end{abstract}

\maketitle

\section{introduction}
%
Measurements of the arrival times of muon neutrinos from the CERN CNGS beam at the OPERA detector 
 730~km away suggest that they travel superluminally with 
$(v-c)/c = (2.37 \pm 0.32 (\rm{stat.}) \pm 0.29 (\rm{sys.})) \times 10^{-5}$~\cite{:2011zb}. Many interpretations of this
stunning result have been proposed~\cite{opera-papers}. 

Inspired by the OPERA anomaly, we present a specific realization of a class of models that may be viewed either as 
superluminal travel of a gauge-singlet sterile neutrino
via extra-dimensional shortcuts~\cite{Pas:2005rb} or alternatively as Lorentz violation for sterile neutrinos as viewed from our 
four-dimensional spacetime~\cite{Coleman:1997xq}. We emphasize at the outset that it is {\it not} our intention to explain
the OPERA data, but to simply provide a concrete model of superluminal neutrinos. 

As described at length in Ref.~\cite{Pas:2005rb},
a superluminal sterile neutrino is well-motivated within the context of 
brane-world phenomenology.
The active neutrinos, carrying electroweak gauge charge like all other Standard Model (SM)
fields, are described as open string excitations with their string endpoints confined to 
our $3+1$ dimensional brane.
On the other hand, the sterile neutrino, carrying no gauge charge, is
characterised by a closed string, free to roam the extra-dimensional bulk as well as the brane
(in the fashion of the ``gauge-singlet'' graviton).
Thus, its geodesic between two points on the brane will include travel in the bulk.
The net result in general will be a shorter transit distance; such shortcuts were proposed a decade ago for
gravitons in Ref.~\cite{Chung:1999zs}.
From the point of view of our brane, the sterile neutrino will appear to travel superluminally.
An analogy would be a comparison of the light transit time and distance when confined 
within a curved optical fiber,
and the light transit time and distance when traveling the straight path between the fiber endpoints.
As proposed in Ref.~\cite{Pas:2005rb},
the shorter distance through the bulk could be a result of brane fluctuations within the bulk.
These fluctuations could be thermal, gravitational, or quantum mechanical in origin.
Also, the difference in the limiting velocities between active and sterile neutrinos $\delta v$ is related to 
the geometry of the brane fluctuation.  The relation $\delta v=(\frac{Ak}{2})^2$ was found,
where $A$ is the (classical) amplitude of the brane fluctuation in the bulk direction,
and $k$ is the wave number of the brane fluctuation along the brane direction.
Thus, $\delta v$ is basically the dimensionless aspect ratio of the brane fluctuation. 

This article is organized as follows. 
We derive the oscillation probabilities when only
one active-sterile mixing angle is nonzero. In doing so, we 
extend the results of Ref.~\cite{Pas:2005rb}, 
from two to three active neutrinos, plus one sterile neutrino.
We then briefly mention aspects of the OPERA data in the context of our model.
Finally, we conclude.

\section{formalism}
%
The quantum mechanics of the model is simple.
The flavor-oscillation amplitude for a propagating neutrino is 
\beq{oscamp1}
A(\nua\rightarrow\nub)=\langle\nub | \,e^{-iHt} |\nua\rangle\,.
\eeq
A component of $Ht$ that is proportional to the identity
cannot affect flavor change, and can be subtracted.  We write the remainder as 
$\delta (Ht)= (\delta H)t+H(\delta t)$ under the assumption that it is small.
We are left with
\beq{oscamp2}
A(\nua\rightarrow\nub)=\langle\nub | \,e^{-i[(\delta H)t+H(\delta t)]} |\nua\rangle\,.
\eeq
As in standard oscillations, $\delta H$ is diagonal in the mass-basis,
and at lowest order is equal to 
\beq{deltaH}
\delta H=\frac{1}{2E}\,{\rm diag}(m^2_1,m^2_2,\cdots)\,.
\eeq
Upon inserting complete sets of mass eigenstates before and after 
$e^{-i(\delta H)t}$ in Eq.~\rf{oscamp2}, the first term there becomes
$\sum_j\,\Uaj^*\,\Ubj\,e^{-i\frac{m^2_j t}{2E}}$;
the usual definition of the bases-mixing matrix,
\beq{Umatrix}
\Uaj=\langle\nua\,|\,\nu_j\rangle, \quad
{\rm or\ equivalently},\ |\nua\rangle=\Uaj^*\,|\nu_j\rangle\,,
\eeq
has been employed.

A nonvanishing value for the second term in Eq.~\rf{oscamp2} is 
unconventional, and occurs if the propagation times for the neutrino states 
are not universal. Such a theory assigns different ``light-cones'' to 
different states, thereby  breaking Lorentz invariance.
Conversely, a large class of models with Lorentz Invariance Violation (LIV) 
has been shown 
to be phenomenologically equivalent to state-dependent limiting 
velocities~\cite{Coleman:1997xq}.
We note that with differing velocities, one has 
$\delta t=\delta(L/v)=-L\,\delta v/v^2$, which is $-L\,\delta v$ to lowest order (in natural units).
In the picture where gauge-singlet states are closed strings free to roam the bulk,
the limiting velocities are assigned to the flavor eigenstates 
rather than to the mass eigenstates.\footnote
{An alternative model arises if one assigns the limiting velocities to the mass eigenstates.
In such a model, the mass-squared matrix and the $\delta v$ matrix are diagonal in the same basis,
and so there is no brane-bulk resonance arising from diagonalization
of the Hamiltonian of Eq.~\rf{Heff1} below. 
Yet another possibility is to assign limiting velocities to velocity eigenstates.
}
Such a choice affects the equivalence between a sterile flavor traveling the shortened geodesics  
available in the bulk, and a Lorentz-violating, superluminal limiting velocity for the sterile state as viewed from the brane.
The second term in \rf{oscamp2} as written is already in a diagonal basis,
and
\beq{deltat}
\delta t={\rm diag}(\delta\ta,\delta\tb,\cdots)=-L\,{\rm diag}(\delta\va,\delta\vb,\cdots)\,.
\eeq
%
%
%

It is conventional to put the physics 
into a Hamiltonian framework.
The effective neutrino Hamiltonian in the flavor basis is
\beq{Heff1}
H_{(F)} =\frac{1}{2E}\,\, U\,
\left(
\barr{cccc}
m^2_1 & 0 & 0 & 0 \\
0 & m^2_2 & 0 & 0 \\
0 & 0 & m^2_3 & 0 \\
0 & 0 & 0 & m^2_4 \\
\earr
\right)
\,U^\dag
-E\,
\left(
\barr{cccc}
\delta v_1 & 0 & 0 & 0 \\
0 & \delta v_2 & 0 & 0 \\
0 & 0 & \delta v_3 & 0 \\
0 & 0 & 0 & \delta v_4 \\
\earr
\right)\,.
\eeq
In general, the $4\times4$ mixing matrix $U$ consists of six angles (the number of planes 
in four dimensions) and four phases. To simplify the analysis, 
we neglect the three new phases, and for now, set to zero the 
rotation angles in the $4-2$  and $4-1$ planes.
By keeping the $\theta_{34}$ angle in $R_{34}$ nonzero, 
we retain the basic features of the model.
We have 
\beq{U4x4}
U=
\left(
\barr{cc}
 V & 0 \\
       0 & 1 \\
\earr
\right)
\times
\left(
\barr{ccc}
 1 & 0 & 0 \\
 0 & 1 & 0 \\
 0 & 0 & R_{34} \\
\earr
\right)\,,
\eeq
in the absence of the new term proportional to $\delta v$'s.
Here, $V$ is the usual PMNS mixing-matrix among the 
three active-flavor neutrinos,
and 
\beq{R34}
R_{34}=
\left(
\barr{cc}
 \cos\theta_{34} & \sin\theta_{34} \\
-\sin\theta_{34} & \cos\theta_{34} \\
\earr
\right)\,.
\eeq
We next write $\Dlsnd\equiv m^2_4 -m^2_3$, and neglect the 
light masses $m^2_j, j=1,2,3$ relative to $m^2_4$.
We assume that the active neutrino flavors have the usual limiting velocity $c$, whereas 
the sterile flavor has a limiting velocity 
$\delta v\equiv \delta v_4>0$.
This seems to us to be the most economic and intuitive 
application of possibly-differing limiting-velocities.
The sterile state is qualitatively different from active states in that it has no 
gauge interactions, and therefore is unconstrained by gauge symmetries.
We provide more discussion of a qualitatively different sterile neutrino below.

With these assumptions, the effective Hamiltonian in~\rf{Heff1} may be written as
\beq{Heff2}
H_{(F)} =
\left(
\barr{cc}
V & 0 \\
0 & 1 \\
\earr
\right)
\left[
\frac{1}{2E}\,
\left(
\barr{ccc}
1 & 0 & 0 \\
0 & 1 & 0 \\
0 & 0 & R_{34} \\
\earr
\right)
\left(
\barr{cccc}
0 & 0 & 0 & 0 \\
0 & 0 & 0 & 0 \\
0 & 0 & 0 & 0 \\
0 & 0 & 0 & \Dlsnd \\
\earr
\right)
\left(
\barr{ccc}
1 & 0 & 0 \\
0 & 1 & 0 \\
0 & 0 & R^T_{34} \\
\earr
\right)
-E\delta v
\left(
\barr{cccc}
0 & 0 & 0 & 0 \\
0 & 0 & 0 & 0 \\
0 & 0 & 0 & 0 \\
0 & 0 & 0 & 1 \\
\earr
\right)
\right]
\left(
\barr{cc}
V^\dag & 0 \\
0 & 1 \\
\earr
\right)\,.
\eeq
The qualitative features of $H_{(F)}$ in Eq.~\rf{Heff2} provide for an 
interesting discussion. 
At sufficiently low energies, the first term on the right-hand-side 
of $H_{(F)}$ dominates,
and oscillations proceed in the standard way.
The second term on the right-hand-side, diagonal 
in the flavor basis, has an analogy with the famous MSW matter-term.  
At sufficiently high energies,
the eigenstates of the Hamiltonian are nearly flavor states, and 
oscillations are very suppressed.
At some intermediate value of energy, the two terms are comparable,
and resonance enhancement of the mixing angles may occur (if the mixing 
angle can reach the maximal-mixing value of $45^\circ$, as discussed below).  

The matrix in brackets in Eq.~\rf{Heff2} is equal to 
\beq{matrix}
\frac{\Dlsnd}{2E}\,
\left(
\barr{cccc}
0 & 0 & 0 & 0 \\
0 & 0 & 0 & 0 \\
0 & 0 & s^2_{34} & s_{34}\,c_{34} \\
0 & 0 & s_{34}\,c_{34} & \left(c^2_{34}-\frac{2E^2\delta v}{\Dlsnd}\right)\\
\earr
\right)\,,
\eeq
and is diagonalized by the rotation $R_{34}$ through an angle 
${\tilde \theta}_{34}$ given by 
\beq{thtilde1}
\tan 2\thtilde=\frac{\sin2\theta_{34}}{\cos2\theta_{34}-2E^2\delta v/\Dlsnd}\,,
\eeq
or equivalently, by 
\beq{thtilde2}
\sin^2 2\thtilde = \frac{\sin^2 2\theta_{34}}{\sin^2 2\theta_{34}
   +\left(\cos2\theta_{34}-2E^2\delta v/\Dlsnd\right)^2}\,.
\eeq
Because of the second term on the right-hand-side of Eq.~\rf{Heff2}, 
we are led to a diagonalization matrix
of the form given in Eqs.~\rf{U4x4} and \rf{R34}, but with $\theta_{34}$ in Eq.~\rf{R34}
replaced by ${\tilde \theta}$.
Thus, the matrix which diagonalizes the full Hamiltonian $H_{(F)}$ is 
\beq{Utilde}
\tU=
\left(
\barr{cc}
 V & 0 \\
       0 & 1 \\
\earr
\right)
\times
\left(
\barr{ccc}
 1 & 0 & 0 \\
 0 & 1 & 0 \\
 0 & 0 & R_{34}(\thtilde) \\
\earr
\right)
= 
\left(
\barr{cccc}
V_{e1}  & V_{e2} & \;V_{e3}\,\cos\thtilde & \;V_{e3}\,\sin\thtilde \\
V_{\mu 1}  & V_{\mu 2} & \;V_{\mu 3}\,\cos\thtilde & \;V_{\mu 3}\,\sin\thtilde \\
V_{\tau 1}  & V_{\tau 2} & \;V_{\tau 3}\,\cos\thtilde & \;V_{\tau 3}\,\sin\thtilde \\
 0 & 0 & -\sin\thtilde & \cos\thtilde \\
\earr
\right)\,.
\eeq

Resonant mixing occurs when the two diagonal elements in Eq.~\rf{matrix} are equal,
{\it i.e.}, when 
\beq{Eres}
\Eres=\sqrt{\frac{\Dlsnd\cos 2\theta_{34}}{2\delta v}}\,.
\eeq
 In terms of $\Eres$, equations~\rf{thtilde1} and \rf{thtilde2} may be written as
\beq{thtilde3}
\tan 2\thtilde=\frac{\tan2\theta_{34}}{1-\left(\frac{E}{\Eres}\right)^2}\,,
\eeq
and
\beq{thtilde4}
\sin^2 2\thtilde = \frac{\sin^2 2\theta_{34}}{\sin^2 2\theta_{34}
   +\cos^2 2\theta_{34}\left(1-\left(\frac{E}{\Eres}\right)^2\right)^2}\,.
\eeq

The energy-dependent angle $\thtilde$ is obtained by taking the inverse sine of Eqs.~\rf{thtilde2} or~\rf{thtilde4},
or the inverse tangent of Eq.~\rf{thtilde1} or~\rf{thtilde3}.
Care must be taken to ensure that $\thtilde$ is chosen in the first octant for $E<\Eres$, 
and in the second octant for $E>\Eres$.
The functions $\sin\thtilde$ and $\cos\thtilde$ are then readily obtained. 
Since $\cos 2\theta_{34}$ is positive definite for small $\theta_{34}$,
resonance can occur only if $\Dlsnd$ and $\delta v$ have the same sign.
Cosmological limits on neutrino masses disallow 
$\sum_{j=1}^3 m_j \ge 3\sqrt{|\Dlsnd|} \sim 3$~eV, 
so $\Dlsnd$ must be positive.  
Thus, resonance is possible only if $\delta v_4 > 0$.
One possibility is to have limiting velocities $v_4=c$, $v_i < c$ ($i<4$).
The other possibility, more natural in the brane-bulk scenario 
and assumed above, is to have $v_i=c$ ($i<4$) and $v_4>c$~\cite{Dent:2007rk}.
This latter possibility is discussed more below.

There are two distinct qualitative differences between the LIV resonance 
inherent in Eq.~\rf{Heff2}, and the MSW matter-resonance.
Firstly, The LIV term here grows with energy, whereas 
the matter term in the MSW Hamiltonian does not.
Consequently, the LIV resonance will be narrower than an MSW resonance.
In other words, a measurement of the full width at half maximum (FWHM) 
may be a signature of the LIV resonance.
Secondly, the LIV resonance does not violate CPT,
whereas the MSW resonance necessarily does; the LIV resonance will occur 
identically in both neutrino and antineutrino channels,
in contrast to the MSW resonance.

The eigenvalues of $H_{(F)}$ are 
\beq{eigvals}
\lambda_1=\lambda_2=0,\quad
\lambda_{4/3}\equiv\lambda_\pm=\frac{\Dlsnd}{4E}\,\left(1-\cos2\theta_{34}\left(\frac{E}{\Eres}\right)^2
\pm\sqrt{\sin^2 2\theta_{34} +\cos^2 2\theta_{34}\,\left[1-\left(\frac{E}{\Eres}\right)^2\right]^2}\,\right)\,,
\eeq
and the eigenvalue differences $\delta H_{kj}\equiv \lambda_k -\lambda_j$ are 
\bea{deltaHkj}
\delta H_{43} &=& \lambda_+ -\lambda_- = \frac{\Dlsnd}{2E}
\sqrt{\sin^2 2\theta_{34} +\cos^2 2\theta_{34}\,\left[1-\left(\frac{E}{\Eres}\right)^2\right]^2}\nonumber\\
\delta H_{42}&=&\delta H_{41}=\lambda_+ \nonumber\\
\delta H_{32}&=&\delta H_{31} = \lambda_- \nonumber\\
\delta H_{21}&=&0\,.
\eea

\noindent

With these eigenvalue differences
and the mixing matrix $\tU$,
we have all the ingredients to obtain all possible oscillation probabilities.\footnote
{In the three-neutrino model of Ref.~\cite{Pas:2005rb}, consisting of two active and one
sterile neutrino, the ``1'' state is absent and so
the $\nu_1$~row and column of $U$ is absent, and $V$ is effectively replaced by the 
$2\times 2$ matrix $R_{23}(\theta_*)$.
Consequently in Ref.~\cite{Pas:2005rb}, $\delta H_{34}$,  $\delta H_{42}$, and $\delta H_{32}$,
are all of the same magnitude in the resonance region.}
Furthermore, in the model as presented, there are just three parameters 
beyond the standard three-neutrino parameters.
These are $\Dlsnd$, $\theta_{34}$, and $\Eres$.{\footnote
{More general LIV scenarios lead to dispersion relations of the form
\begin{equation}
\label{disprel}
E\sim |\vec{p}| + \frac{m^2}{2 |\vec{p}|} \pm \delta v \frac{|\vec{p}|^n}{E_0^{n-1}},
\end{equation}
where $E_0$ denotes some typical energy scale. The case under discussion correponds to $n=1$.
For arbitrary $n$,
\begin{equation}\label{Eres}
\Eres = \bigg({\frac{\Delta\,E_0^{n-1}\cos 2\theta_{34}}{2\eps}}\bigg)^{1\over{n+1}}\,,
\end{equation}
and the corresponding $\sin^2 2\thtilde$ and $\delta H_{kj}$ are obtained by replacing
$(E/\Eres)^2$ by $(E/\Eres)^{n+1}$ in Eqs.~\rf{thtilde3}-\rf{deltaHkj}.
}

The general oscillation formulae are
\beq{oscprob1}
P(\nua\rightarrow\nub)=\delta_{\alpha\beta}
-4\sum_{j<k} \Re\{\tildeUbj\,\tildeUbk^*\,\tildeUaj^*\,\tildeUak\}\,\sin^2\left(\frac{L\,\delta H_{kj}}{2}\right)
   +2 \sum_{j<k} \Im\{\tildeUbj\,\tildeUbk^*\,\tildeUaj^*\,\tildeUak\}\,\sin\left(L\,\delta H_{kj}\right) \,,
\eeq
%
which on ignoring phases in $U$ becomes 
\beq{oscprob1.5}
P(\nua\rightarrow\nub) = \delta_{\alpha\beta}
   -4\sum_{j<k} \tildeUbj\,\tildeUbk\,\tildeUaj\,\tildeUak\,\sin^2\left(\frac{L\,\delta H_{kj}}{2}\right)\,.
\eeq
For the present case we get
\beq{oscprob2}
P(\nua\rightarrow\nub) =\delta_{\alpha\beta}-4\times
\left\{
\barr{ll}   
  \sin^2\left(\frac{L\,(\lambda_+ -\lambda_-)}{2}\right)&\tU_{\beta 3}\,\tU_{\beta 4}\,\tU_{\alpha 3}\,\tU_{\alpha 4}\\
  +\sin^2\left(\frac{L\,\lambda_+ }{2}\right)            &\sum_{j=1,2}\,\tU_{\beta j}\,\tU_{\beta 4}\,\tU_{\alpha j}\,\tU_{\alpha 4}\\
  +\sin^2\left(\frac{L\,\lambda_- }{2}\right)            &\sum_{j=1,2}\,\tU_{\beta j}\,\tU_{\beta 3}\,\tU_{\alpha j}\,\tU_{\alpha 3}\,.
\earr
\right.
\eeq

A relevant variable for neutrino oscillations is $L/E$.
Just as $\Eres$ sets the energy scale for the resonance, 
the length scale for the $n$-th maximum at resonance is set by an interplay of the various 
\beq{Lres1}
\Lres\equiv \frac{\pi\,(2n-1)}{|\delta H_{jk}|}\,.
\eeq
Substituting $E=\Eres$ into Eqs.~\rf{eigvals} and~\rf{deltaHkj}, we find 
the values of $L/E$ at $E=\Eres$ for the three contributing amplitudes to be
\beq{Lres2}
\left(\frac{\Lres^{+-}}{\Eres}\right) \equiv
\frac{\pi\,(2n-1)}{\Eres\,(\lambda_+ -\lambda_-)}=\frac{2\pi\,(2n-1)}{\Delta\,\sin2\theta_{34}}\,,
\eeq
and
\beq{Lres3}
\left(\frac{\Lres^\pm}{\Eres}\right) \equiv
\frac{\pi\,(2n-1)}{\Eres\,|\lambda_{\pm}|}=\frac{4\pi\,(2n-1)}{\Delta\,(\sin2\theta_{34}\pm2\sin^2\theta_{34})}
\approx 2\,\left( \frac{\Lres^{+-}}{\Eres} \right)\,.
\eeq

Note that the $1-2$ submatrix of $\tU$ is the same as that of $V$.
The matrix $V$, like $\tU$, is unitary.
Thus, $\sum_{j=1,2} \tildeUaj\,\tildeUbj = \delta_{\alpha\beta} -V_{\alpha 3}\,V_{\beta 3}$.
Making this replacement, and using the explicit matrix entries in the third and 
fourth columns of Eq.~\rf{Utilde},
we arrive at simpler expressions for the three relevant cases: active neutrino survival, 
active-to-active neutrino conversion, and active-to-sterile conversion.
Denoting the sterile neutrino by $\nu_s$ and active flavors by $\nu_a, \nu_b,\cdots$, 
the active neutrino survival probability is given by 
\bea{active_survival}
P(\nu_a\rightarrow\nu_a) &=&1-4\,V^2_{a3}\times
\left\{
\barr{ll}   
  \sin^2\left(\frac{L\,(\lambda_+ -\lambda_-)}{2}\right)&\sin^2\thtilde\,\cos^2\thtilde\;\;V^2_{a3}\\
  +\sin^2\left(\frac{L\,\lambda_+ }{2}\right)            &\sin^2\thtilde\;\;(1-V^2_{a3})\\
  +\sin^2\left(\frac{L\,\lambda_- }{2}\right)            &\cos^2\thtilde\;\;(1-V^2_{a3})\,.
\earr
\right.
\eea
The active-to-(different) active neutrino conversion probability is given by
(and mind the minus sign on the first term in brackets)
\bea{active_conversion}
P(\nu_a\rightarrow\nu_b) &=& 4\,V^2_{a3}\,V^2_{b3}\times
\left\{
\barr{ll}   
  -\sin^2\left(\frac{L\,(\lambda_+ -\lambda_-)}{2}\right)&\sin^2\thtilde\,\cos^2\thtilde\\
  +\sin^2\left(\frac{L\,\lambda_+ }{2}\right)            &\sin^2\thtilde\\
  +\sin^2\left(\frac{L\,\lambda_- }{2}\right)            &\cos^2\thtilde\,.
\earr
\right.
\eea
The active-to-sterile conversion probability is given by
\beq{active_sterile}
P(\nu_a\rightarrow\nu_s)=V^2_{a3}\,\sin^2 2\thtilde\,\sin^2\left(\frac{L\,(\lambda_+ -\lambda_-)}{2}\right)\,.
\eeq

Note that correct limits are respected here.
Far above the resonance, $\cos^2\thtilde$ and $\lambda_+$ approach zero
(while $\sin^2\thtilde$ and $\lambda_-$ do not).
Thus, each term in the above probabilities vanishes far above $\Eres$,
and the sterile state effectively decouples, as it must.

The analytic formalism presented here fails if more than one $\theta_{j4}$ is taken to be nonzero,
for then the eigenvalues must be derived from a matrix larger than the 
$2\times 2$ subblock given in Eq.~\rf{matrix}.
However, the formalism goes through when a $\theta_{j4}$ other than
$\theta_{34}$ is taken to be nonzero.  
We have really described three models here, 
characterized by a nonzero $\theta_{34}$, $\theta_{24}$, or $\theta_{14}$. 
In Eqs.~\rf{active_survival}--\rf{active_sterile},
one need only replace the subscript ``3'' by ``2'' or ``1'' to 
obtain the $\theta_{24}$ and $\theta_{14}$ models, respectively. 

For the $\theta_{34}$ model, we have 
$4\,V^2_{e3}\,V^2_{\mu 3}= \sin^2 (2\theta_{13})\,\sin^2 \theta_{23}$,
$4\,V^2_{e3}\,V^2_{\tau 3}= \sin^2 (2\theta_{13})\,\cos^2 \theta_{23}$,
and
$4\,V^2_{\mu 3}\,V^2_{\tau 3}= \sin^2 (2\theta_{23})\,\cos^4 \theta_{13}$,
for the prefactors to $P(\nue\leftrightarrow\numu)$,
$P(\nue\leftrightarrow\nutau)$,
and $P(\numu\leftrightarrow\nutau)$, respectively.

The formalism needs to be extended to be relevant for data away from the resonance. 
Continuing with the simple model with just one nonzero $R_{j4}$, we see that
a factorization occurs between the squared elements of $V$ and $R$ in $U$, {\it viz}. 
(with no sum on $j$ implied)
\bea{factorize}
U^2_{\alpha k} &=& [ \sum_p V_{\alpha p}\,(R_{j4})_{pk}]^2
   =\sum_{p,q} V_{\alpha p}\,(R_{j4})_{pk}\,V_{\alpha q}\,(R_{j4})_{qk}= [ V_{\alpha j} ]^2\, [(R_{j4})_{jk} ]^2 \,.
\eea
The final form results because only the $j^{th}$ active flavor and the sterile 
state appear in the $R_{j4}$ matrix. 
This result is simple matrix multiplication of the matrix with $V$-squared 
elements and the matrix with $R$-squared elements.
It is useful to present the squared elements of the $V$ and $R$ matrices.
The $R$-squared elements are 
\beq{Rsquared}
\left(
\barr{cccc}
 1 & 0 & 0 & 0 \\
 0 & 1 & 0 & 0 \\
 0 & 0 & \cos^2\thtilde & \sin^2\thtilde \\
 0 & 0 & \sin^2\thtilde & \cos^2\thtilde \\
\earr
\right)\,,
\eeq
with obvious generalizations for the other choices of nonzero $R_{4j}$.
We next discuss the $V$-squared matrix.

A PMNS matrix consistent with most neutrino data 
is the tribimaximal matrix~\cite{Harrison:2002er}, which extended to the $4\times4$ case is:
\beq{tribimax1}
V_{4\times 4}=\frac{1}{\sqrt{6}}\,
\left(
\barr{rrrr}
2  & \sqrt{2}  & 0        & 0 \\
-1 & \sqrt{2} &  \sqrt{3} & 0 \\
-1 & \sqrt{2} & -\sqrt{3} & 0 \\
0 & 0 & 0 & \sqrt{6} \\
\earr
\right)\,.
\eeq
The $V$-squared elements of this matrix are
\beq{tribisquared}
\frac{1}{6}\,
\left(
\barr{cccc}
 4 & 2 & 0 & 0 \\
 1 & 2 & 3 & 0 \\
 1 & 2 & 3 & 0 \\
 0 & 0 & 0 & 6 \\
\earr
\right)\,.
\eeq
For example, in the $\theta_{34}$-model, 
$U^2_{\mu 3}=[V_{\mu 3}]^2\,[(R_{34})_{33}]^2= \half\cos^2\thtilde$.
This can also be seen from Eq.~\rf{Utilde}.
A unitary extension of the tribimaximal matrix to nonzero $U_{e3}$ is given in 
Ref.~\cite{triminimal}.
%
%
We find that the conversion probability for $\numu\rar\nue$, to lowest order in $|U_{e3}|^2$, is
\beq{mu2e_conversion}
P(\nu_\mu\rightarrow\nu_e) =
\left\{
\barr{ll}
    4/9 & \ {\rm for}\ j=1,2\\
    2\,|U_{e3}|^2 & \ {\rm for}\ j=3
\earr
\right\}
\times
\left\{
\barr{ll}   
  -\sin^2\left(\frac{L\,(\lambda_+ -\lambda_-)}{2}\right)&\sin^2\thtilde\,\cos^2\thtilde\\
  +\sin^2\left(\frac{L\,\lambda_+ }{2}\right)            &\sin^2\thtilde\\
  +\sin^2\left(\frac{L\,\lambda_- }{2}\right)            &\cos^2\thtilde\,.
\earr
\right.
\eeq
Note that the different choices of the nonzero $\theta_{j4}$ result in 
just a change in the overall magnitude of the flavor-changing probability.

%
To study the influence of the new mass-squared scale $\Dlsnd$ on 
 long-baseline and atmospheric data, the atmospheric mass-squared 
scale $\Datm\equiv |m^2_3 -m^2_1|$ 
must be entered into the probability formulae.
This is done simply in the following way:
In Eq.~\rf{Heff2}, when $\theta_{34}$ is chosen for mixing
the diagonal mass matrix is replaced by 
\beq{atmos_scale1}
\left(
\barr{cccc}
0 & 0 & 0 & 0 \\
0 & 0 & 0 & 0 \\
0 & 0 & 0 & 0 \\
0 & 0 & 0 & \Dlsnd \\
\earr
\right)
\rightarrow
\left(
\barr{cccc}
\mp\Datm &     0    & 0 & 0 \\
    0    & \mp\Datm & 0 & 0 \\
0 & 0 & 0 & 0 \\
0 & 0 & 0 & \Dlsnd \\
\earr
\right)
\eeq
and when $\theta_{24}$ or $\theta_{14}$ are chosen for mixing, 
the diagonal mass matrix is replaced by 
\beq{atmos_scale2}
\left(
\barr{cccc}
0 & 0 & 0 & 0 \\
0 & 0 & 0 & 0 \\
0 & 0 & 0 & 0 \\
0 & 0 & 0 & \Dlsnd \\
\earr
\right)
\rightarrow
\left(
\barr{cccc}
0 & 0 & 0 & 0 \\
0 & 0 & 0 & 0 \\
0 & 0 & \pm\Datm & 0 \\
0 & 0 & 0 & \Dlsnd \\
\earr
\right)
\eeq
These choices ensure that $\Datm$ is not mixed into the resonance condition,
making the extraction of the four eigenvalues simple. 
We neglect effects of order $\Datm/\Dlsnd$.
The two sign choices correspond to normal and inverted hierarchies, respectively, 
for the active mass spectrum. 
Notice that because one of the three active states is distinguished from the other two,
due to its mixing with $\nu_4$, the two hierarchies yield different 
physics.  This is analogous to the the MSW situation in matter.

For the $\theta_{34}$ model, the ordered eigenvalues of the full Hamiltonian are 
$\lambda_j=\{\mp\Datm/2E,\mp\Datm/2E,\lambda_-,\lambda_+\}$;
there are three distinct nonzero eigenvalue differences $\delta H_{kj}=\lambda_k-\lambda_j$,
as in the $\Datm=0$ case discussed earlier.
To get the oscillation probabilities, 
these eigenvalue differences $\delta H_{kj}=\lambda_k-\lambda_j$ are 
inserted into Eq.~\rf{oscprob1.5}.

For the $\theta_{24}$ and $\theta_{14}$ models, the eigenvalues are 
$\lambda_j=\{0$ and $\lambda_-$ in the appropriate order, followed by $\pm\Datm/2E$ and $\lambda_+\}$;
here there are six distinct eigenvalue differences.
After insertion into Eq.~\rf{oscprob1.5}, one obtains the oscillation probabilities 
after some straightforward but tedious algebra.

\section{Opera}

OPERA has inferred a mean arrival time for muon neutrinos of $\sim 60$~ns faster than the theoretical light-travel time,
over the 730~km pathlength.
The short-bunch (3 ns) beam data newly acquired by OPERA provide a stringent constraint on models in which 
neutrino flavor states or mass states travel with different limiting velocities.
This is because the new data show an arrival time rms of $\sim 16$~ns about the mean, 
with {\it no events} arriving within 37~ns of the theoretical light-travel time.
It appears that there is but one neutrino speed in the data.
A fit to these new data~\cite{Winter:2011zf} 
requires the fraction of superluminal neutrinos to be at least 80\% at a $3\sigma$ confidence. 
(Resonant conversion offers the possibility of a 50\% mixture on average, 
while adiabatic conversion and back-conversion can theoretically attain 100\% conversion to the fast species.)
This implies that atmospheric neutrinos oscillate primarily into sterile neutrinos, a conclusion that is excluded
by Super-Kamiokande.

Another daunting constraint 
comes from the $L/E$ distribution of Super-Kamiokande's atmospheric muon neutrino events~\cite{SKatmos}.
A large scale (compared to $2.5\times 10^{-3}$~eV$^2$) in the difference of eigenvalues of the Hamiltonian, as occurs in active-plus-sterile neutrino models (cf. Eqs.~\rf{eigvals}, \rf{deltaHkj} and \rf{active_survival})
would provide an energy-independent, averaged contribution for the $L/E$ distribution,
in violation of these data, unless the active-sterile mixing angle is small.
However, the OPERA data tell us that the mixing in an overlapping $L/E$ range cannot be small.
Thus, the simple model we have presented does not simultaneously account for OPERA and 
atmospheric data.

On the positive side, additional constraints on models of OPERA data seem to be easy to meet 
with our class of models.
The coincident time of arrival of neutrinos and photons from SN 1987A is easily accommodated 
by making $\theta_{13}$ sufficiently small so as to suppress  
$\nu_e \to \nu_s$ oscillations; see Eq.~\rf{active_sterile}.
Also, the analogue of 
Cherenkov radiation for superluminal neutrinos, pointed out in Ref.~\cite{Cohen:2011hx}, does not
apply at least to the extra-dimensional variety
of our model since all propagation is subluminal locally, and the apparent superluminal behavior
 is simply a consequence of the bulk shortcut.
(Also, with SM particles confined to the brane, 
there can be no Cherenkov radiation of SM particles from sterile neutrinos traveling in the bulk.)

\section{Conclusions}

In light of the alleged evidence for superluminal propagation of neutrinos in OPERA data, and drawing
on our prior work on superluminal neutrinos, 
we have provided herein a concrete model that accommodates superluminal neutrinos.
We find that oscillations between active neutrinos and a sterile neutrino 
 with a shorter geodesic path through extra dimensions
(or equivalently, obeying a modified dispersion relation as seen from the brane) 
do not provide a simple explanation of the OPERA anomaly because of conflicts with 
atmospheric neutrino data.  This does not invalidate the model in a broader context.
Introducing additional sterile neutrinos and/or mixing angles into the framework may
produce consistency with data, but we have not explored this possibility. 
In one variation of the model, the resonance may have an $(LE)$-dependence~\cite{Hollenberg:2009ws} 
(instead of a simple $E$-dependence) 
thus affording greater flexibility in addressing the tension with atmospheric data.
It does seem, however, that an explanation of the OPERA anomaly using neutrino oscillations
is likely to be contrived. 

 Finally we remark on the generality of our results.
Since the bulk-shortcut model for sterile neutrinos appears from the vantage point of the brane as a LIV model,
our conclusions apply in generality to the larger class of models in which a sterile flavor state 
is superluminal, and transmits its greater speed to active flavor states via mixing and oscillations.

\vspace{0.5cm}
{\it{Acknowledgments.}}
We thank J.~Kumar and J.~Learned for discussions. 
DM thanks the University of Hawaii for its hospitality while
this work was in progress. HP thanks the University of Hawaii and the Universidad Tecnica Federico Santa Maria,
Valparaiso, Chile for their hospitality.
This research was supported by US DoE Grants DE-FG02-04ER41308, 
DE-FG05-85ER40226 and DE-FG02-04ER41291,
by US NSF Grant PHY-0544278 and by DFG Grant No. PA 803/5-1.

\end{document}